\documentclass[journal=jacsat,manuscript=letter]{achemso}
\setkeys{acs}{articletitle = true}

\usepackage{chemformula} 
\usepackage[T1]{fontenc} 
\usepackage{amssymb}

\author{Wushi Dong}
\affiliation[The University of Chicago]
{James Franck Institute and Department of Physics, The University of Chicago, Chicago, Illinois 60637, United States}
\email{dongws@uchicago.edu}

\author{Peter B. Littlewood}
\affiliation[The University of Chicago]
{James Franck Institute and Department of Physics, The University of Chicago, Chicago, Illinois 60637, United States}
\alsoaffiliation[Argonne National Laboratory]
{Materials Science Division, Argonne National Laboratory, Lemont, Illinois 60439, United States}

\title[]{Quantum electron transport in ohmic edge contacts between two-dimensional materials}

\abbreviations{}

\begin{document}


%
%
%



\newpage
\begin{abstract}
The metal-semiconductor contact is a major factor limiting the shrinking of transistor dimension to further increase device performance.
In-plane edge contacts have the potential to achieve lower contact resistance due to stronger orbital hybridization compared to conventional top contacts.
However, a quantitative understanding of the electron transport properties in the edge contact is still lacking. In this work, we present full-band atomistic quantum transport simulations of the graphene/\ch{MoS2} edge contact. By using a Wannier function basis to accurately describe the electronic bands, together with a full self-consistent solution of the electrostatics, we are able to efficiently model device structures on a micron scale, but with atomic level accuracy.
We find that the potential barrier created by trapped charges decays fast with distance away from the interface, and is thus thin enough to enable efficient injection of electrons. This results in Ohmic behavior in its I-V characteristics, which agrees with experiments.
Our results demonstrate the role played by trapped charges in the formation of a Schottky barrier, and how one can reduce the Schottky barrier height (SBH) by adjusting the relevant parameters of the edge contact system.
Our framework can be extended conveniently to incorporate more general nanostructure geometries. For example, a full 3D solution of the electrostatics will also lead to better modeling of the electrical potential. Furthermore, better ab-initio calculations can be conveniently added to our methods to further improve their accuracy.
\end{abstract}
\newpage

Two-dimensional (2D) materials such as graphene and transition metal dichalcogenides (TMDCs) are pushing the forefront of complementary metal-oxide semiconductor (CMOS) technology beyond the Moore's law \cite{neto2009electronic, novoselov20162d}, and show great promises for realizing atomically thin circuitry \cite{levendorf2012graphene, fiori2014electronics, duan2015two}.
A fundamental challenge to their effective use remains the large resistance of electrical contacts to 2D materials for probing and harnessing their novel electronic properties \cite{allain2015electrical, xu2016contacts, schulman2018contact}.
There are generally two types of contact geometries, namely top contacts and edge contacts \cite{allain2015electrical}.
Conventional 3D metallic top contacts can achieve low contact resistance with monolayer 2D materials, but cannot avoid the intrinsic problem of large electrode volume.
\cite{li2014thickness, liu2015impact, allain2015electrical, liu2018approaching}
2D top contacts, including graphene \cite{das2012high, das2014all, liu2015toward} and recently demonstrated atomically flat metal thin films \cite{liu2018approaching}, can achieve both small volumes and low contact resistances of metal-semiconductor interfaces, but they suffer from weak van der Waals coupling to TMDCs \cite{kang2014computational}. Their transfer efficiency depends largely on the contact area and is compromised dramatically below a transfer length which is typically tens of nm scale \cite{ling2016parallel, schulman2018contact}.
In contrast, 2D edge contacts are formed by joining atomically thin metal electrodes and semiconductors laterally in a single plane. They offer the possibilities for high-quality contacts to 2D materials despite minimal contact area defined by their atomic thickness as shown by both simulations \cite{kang2014computational} and recent experimental successes \cite{ling2016parallel, guimaraes2016atomically}.
Among them, the graphene-\ch{MoS2} system considered in this paper is particularly promising for a low-resistance 2D edge contact \cite{ling2016parallel, zhao2016large, guimaraes2016atomically}. According to the Schottky-Mott rule, the combination of a low-work-function metallic graphene electrode and a typical n-type \cite{lee2012synthesis} semiconducting monolayer \ch{MoS2} channel naturally leads to a small SBH. Moreover, the overall system is stable under working conditions and resistant to phase transitions induced by adsorbates.
While improving experimental techniques makes more tests feasible, a better quantitative understanding of the electronic structure and transport properties is still critical for improving the design of 2D edge contacts.
In \citeyear{yu2016carrier}, \citeauthor{yu2016carrier} \cite{yu2016carrier} suggested a highly non-localized carrier redistribution and strong reduction of Fermi level pinning in 2D systems based on a semi-classical macroscopic model. In \citeyear{chen2017properties}, \citeauthor{chen2017properties} \cite{chen2017properties} performed first-principles studies based on density functional theory (DFT) on the morphologies of the graphene/\ch{MoS2} lateral junction and proposed several stable interface configurations. \citeauthor{sun2017first} \cite{sun2017first} performed similar studies and tried to calculate the transport efficiencies, but did not reproduce the linear I-V characteristics observed in experiments, possibly due to the lack of doping in the semiconductor region.

In order to better model the graphene/\ch{MoS2} interface at the atomic scale and quantitatively calculate the charge transfer properties, we introduced a custom-built self-consistent quantum transport solver based on the Keldysh Nonequilibrium Green's Function formalism \cite{luisier2006atomistic, datta2000nanoscale} and Maximally Localized Wannier functions (MLWFs) \cite{marzari2012maximally}. Such a method can efficiently solve the local electrostatics and electron transport with first-principles accuracy at a minimal cost of tight-binding calculations.
It enables the inclusion of large areas of both materials, which is necessary in order to allow for a long screening length for charged interfacial states and thus to have equilibrium conditions near the edge of the central device region. 
This is also a necessary condition for the decimation technique to account for the effects of the semi-infinite leads.
We find that trapped interface states lead to a potential barrier, which is however small enough that we find Ohmic behavior at room temperature and high enough doping levels. We successfully reproduced the linear current-voltage (I-V) characteristics with a resistivity value of approximately 30 $k\Omega \cdot \mu m$ close to that observed in experiments \cite{guimaraes2016atomically} at room temperature, which is a first for 2D edge contact systems. At lower temperatures, We observe increasing non-linearity as a result of reduced thermalization.
In the following, we calculate the band structures of graphene and monolayer \ch{MoS2}, and extract their tight-binding parameters using MLWFs. Based on these parameters, we use our custom-built quantum transport solver to generate the electrostatic potential self-consistently with the inhomogeneous charge densities induced by band bending, and by local impurity states. We confirm the validity of our solver by comparing the converged electrostatic potential profile to the analytical predictions from Thomas-Fermi screening theory, beyond angstrom distances from the contact region. Finally, we calculate the transport properties based on the Keldysh formalism and discuss how to further improve device performance.

\begin{figure}
\includegraphics[width=\linewidth]{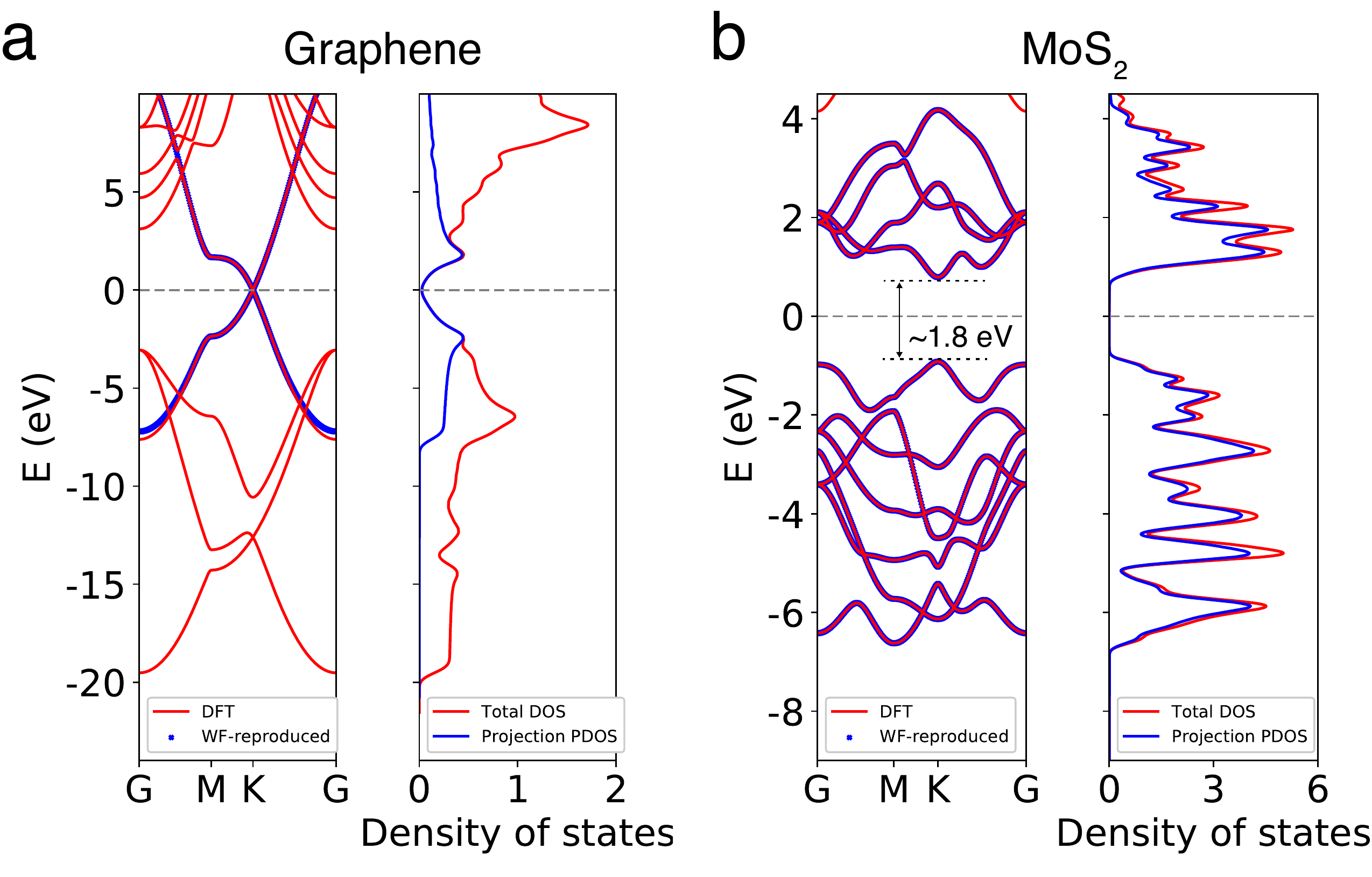}
  \caption{(a) Left: Graphene band structures obtained with DFT and with MLWF Hamiltonian. Right: Graphene total DOS and its PDOS reproduced by the MLWF orbitals. (b) The same plots for monolayer \ch{MoS2}. The produced band gap of approximately 1.8 $eV$ is very close to the experimental one. The zero energy is set to the Fermi level (dashed line) for both materials.}
  \label{bands}
\end{figure}

Wannier functions can accurately and efficiently capture delicate electronic structures. 
We used them to extract the tight-binding parameters of both materials after obtaining the band structures in the DFT framework.
Figure \ref{bands} compares the band structures obtained with DFT and with the MLWF Hamiltonian for both monolayer graphene and \ch{MoS2}. From the plot, we can see that the Wannier projections work so well that the differences between the two bands are largely unnoticeable.
We also compare the total Density of States (DOS) and the Projected Density of States (PDOS) reproduced by the MLWF orbitals for both materials. Instead of using all the orbitals in a unit cell, we choose only those contributing to the DOS near the Fermi level for Wannier projections. This can minimize the sizes of matrices used in our calculations and further reduce computational cost.
The extracted hopping parameters then serve as basis for the transport simulations.

\begin{figure}[H]
\includegraphics[width=\linewidth]{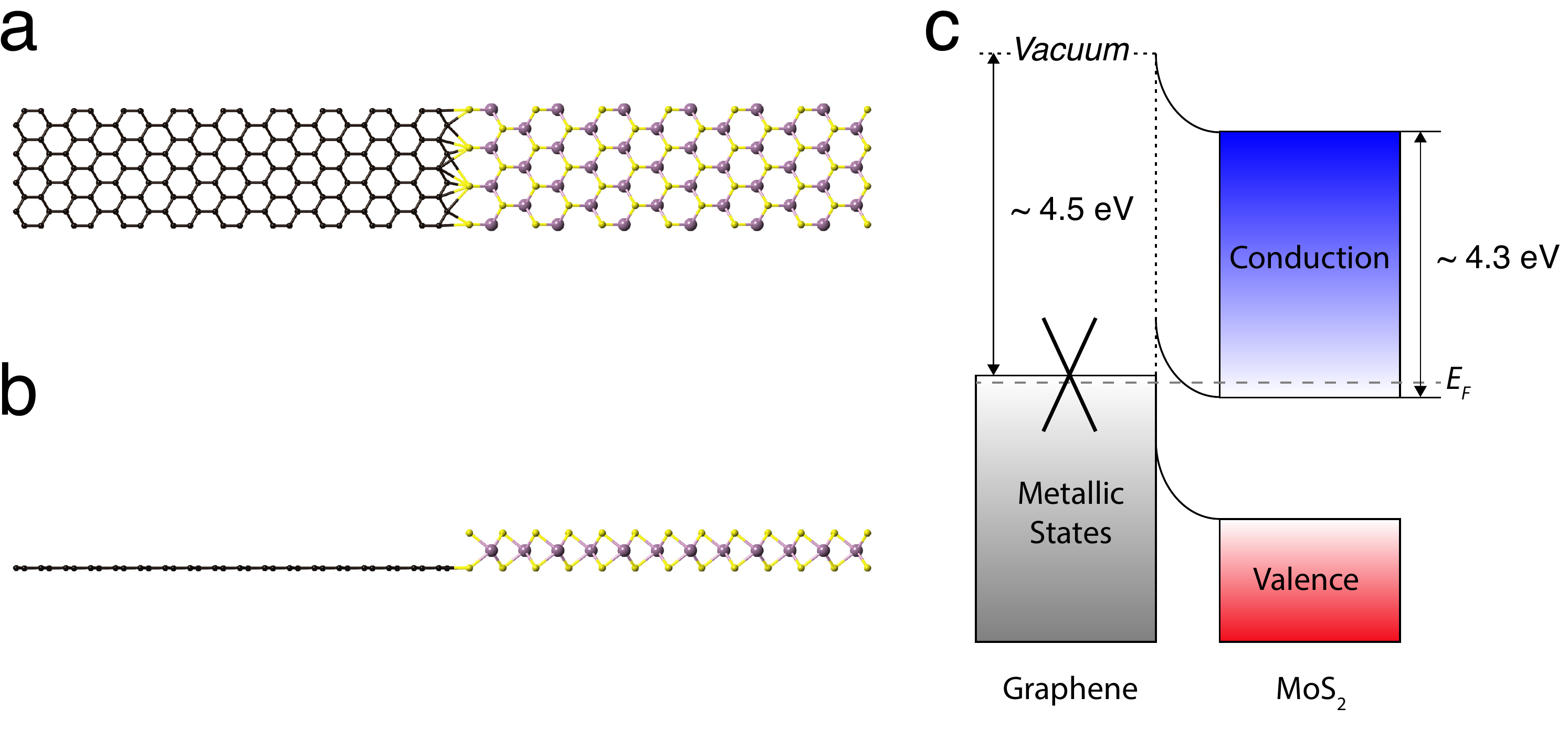}
  \caption{(a) Top, and (b) side views of the simulated edge contact device region. (c) Schematic illustration of its band alignments. The work function of graphene is about 4.5 $eV$ and the electron affinity of monolayer \ch{MoS2} is 4.3 $eV$ \cite{chen2017properties}. $E_F$ stands for Fermi energy.}
  \label{geometry}
\end{figure}

The geometry of the graphene-\ch{MoS2} edge contact is sketched in Figure \ref{geometry}a and \ref{geometry}b. We assume periodicity of the device in the y direction, at a level of a few unit cells which is long enough to approximately match the lattice constants. As a result, the Hamiltonian shows a $k_y$ dependence and can be decomposed into three components according to the Bloch's theorem as
\begin{equation}
H(k_y) = H_0 + H_{-} e^{-\mathrm{i}k_y\Delta} + H_{+} e^{\mathrm{i}k_y\Delta},
\end{equation}
where $H_0$ are the interactions within a strip of width, $H_{\pm}$ are the interactions with a neighbor strip along the $+y$ or $-y$ direction, and $\Delta$ is the width of the supercell. To make second-nearest-neighbor interactions negligible, we choose $\Delta$ to be exactly $4$ times the width of the graphene unit cell, and approximately $3$ times that of the \ch{MoS2} unit cell, resulting in a lattice mismatch of only 4.2\%. We do not change the lattice constant of graphene because it has a much larger Young's modulus \cite{jiang2015graphene}.
For the interface, we consider the predominant zigzag edge of graphene \cite{yu2011control} and \ch{MoS2} \cite{lauritsen2007size} as shown in Figure \ref{geometry}a and choose a structure motivated by {\em ab initio} calculations. According to \citeauthor{chen2017properties} \cite{chen2017properties}, the configuration chosen in our study has the lowest formation energies among other alternative geometries.
We adjust the Fermi level of both materials to match the induced doping by gate voltage. 
For the details of the parameters used in this study and their effects on the simulation results, please refer to our supporting information.
Figure \ref{geometry}c illustrates the band alignments of the edge contact device simulated in this paper.
Having established the atomic geometry and band alignments of the junction, we now investigate the electrostatics and charge transfer effects at the boundary.

\begin{figure}
\includegraphics[width=\linewidth]{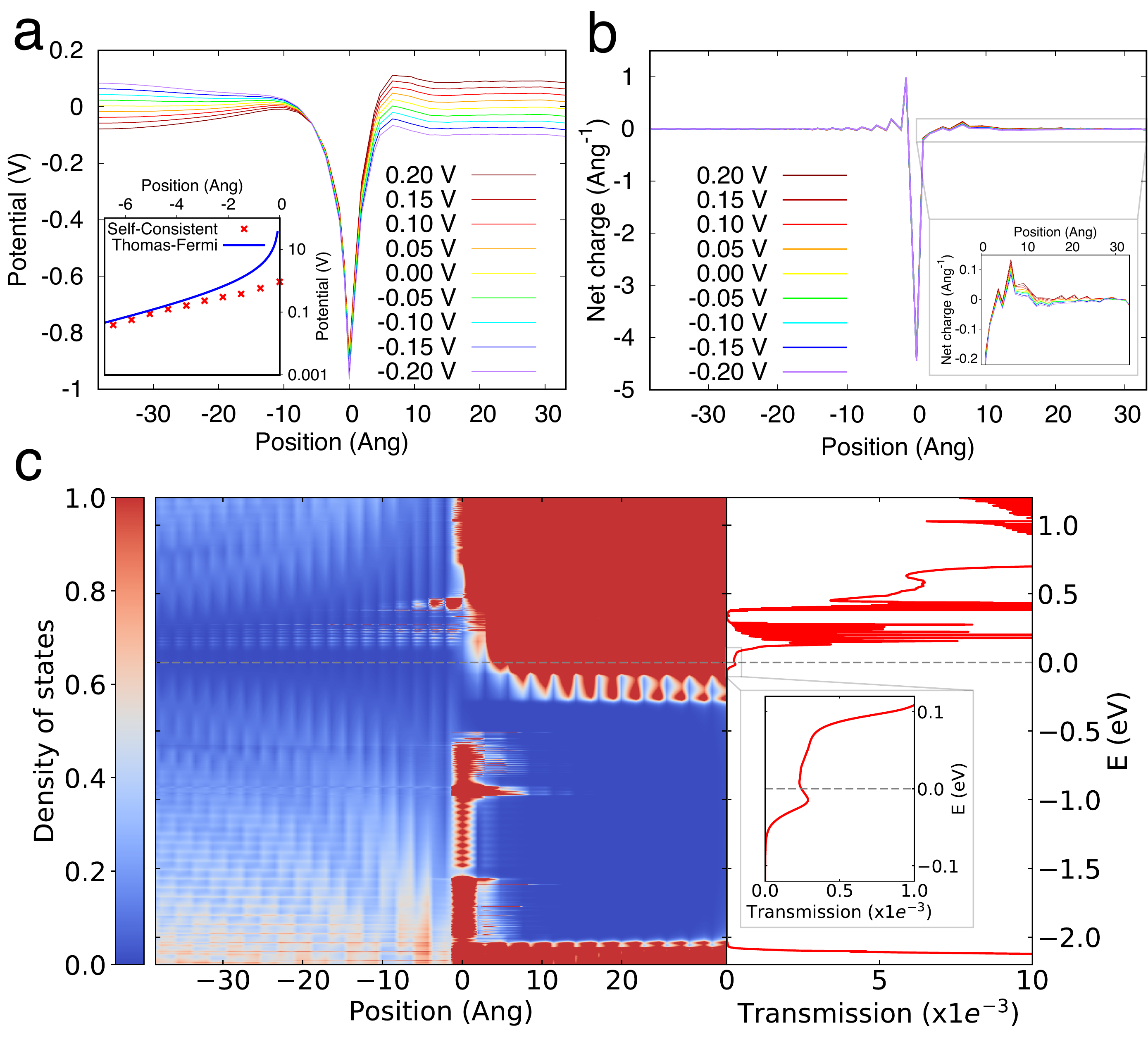}
  \caption{(a) Converged electrostatic potential profiles under different biases. Inset: Comparison between the results of 1D Thomas-Fermi screening (solid line) and our self-consistent simulation (dots) on a log scale at zero bias. (b) Converged net charge profiles under different biases. Inset: The same results enlarged for the \ch{MoS2} side to show the difference under different biases. (c) LDOS and transmission spectrum of the simulated graphene-\ch{MoS2} edge contact at zero bias. Inset: Transmission spectrum near the Fermi level at $E = 0$.}
  \label{electrostatics}
\end{figure}

To evaluate the tunneling barrier, we calculate charge densities self-consistently with electrostatic potential for the edge contact device. We use the Nonequilibrium Green's Function technique based on the Keldysh formalism to calculate the charge densities, and the non-linear Newton-Raphson technique to solve for the electrostatic potential from the Poisson equation.
We performed simulations for different source-drain biases at a {MoS2} doping level of $4\times10^{14} cm^{-2}$ and at room temperature of $T = 293K$. The converged potential and charge profiles are shown in Figure \ref{electrostatics}a and Figure \ref{electrostatics}b respectively.
We can see that the electrostatic potential reaches equilibrium at both edges of our simulated region. Applied source-drain biases shift the potential level in the two electrodes and in turn modify the net charge profile. In the inset of Figure \ref{electrostatics}b, we enlarge the plot for the \ch{MoS2} side to better show how the net charge distribution adapts to the external biases.
Here we safely ignored the electron-phonon scattering, due to the short channel length of 2D edge contacts.
In order to check the validity of the electrostatics obtained from our self-consistent simulations, we compare the converged potential profile with the analytical predictions of a quasi-1D Thomas-Fermi screening potential, and the result is shown in the inset of Figure \ref{electrostatics}a. We find that the two results agree well beyond angstrom distances from the contact region, where the charge densities can stay low enough for the Thomas-Fermi theory to work well. This confirms the accuracy of our method. The derivations of the quasi-1D Thomas-Fermi screening potential are given in the supporting information.

Using the converged electrostatic profiles, we further examine the quantum transport properties of the graphene-\ch{MoS2} edge contact by calculating its Local Density of States (LDOS) and transmission spectrum using the Landauer-Buttiker formalism.
The transmission coefficients are determined by the equation:
\begin{equation}
  T = Tr[G^R\Gamma_LG^A\Gamma_R]
  \label{transmission}
\end{equation}
where $\Gamma_L$ and $\Gamma_R$ are the linewidth functions that describe the coupling between the scattering region and the two leads on the left and right.
In Figure \ref{electrostatics}c, We show the two results under room temperature and zero bias side by side.

From the results in Figure \ref{electrostatics}, we find that the electrostatic potential is screened by the electrons and decays fast as one goes away from the interface, which allows charge carriers to tunnel through the boundary efficiently.
When the metallic graphene contacts with monolayer \ch{MoS2} contacts, free electrons will flow from the graphene side to the \ch{MoS2} side since the work function of p-type graphene is smaller than that of n-type \ch{MoS2}.
When the charge redistribution reaches equilibrium, graphene is positively charged whereas \ch{MoS2} monolayer is negatively charged near the interface region, in which a built-in electric dipole is induced. \cite{bristowe2014origin}
In addition, trapped charges at the interface produce a monopole, which we find to be substantial. Such electric fields can shift the energy bands of the \ch{MoS2} monolayer upward. 
However, we find that the barrier is efficiently screened by the free charges and becomes thin enough for the electrons to tunnel through. 
From Figure \ref{electrostatics}c, we can see the two-fold effects of the interfacial bonding: The trapped charges at the interface form a thin potential barrier, which is screened effectively allows electrons to go through; The overlap states serve as a bridge inside the barrier further assisting with charge transfer.
As a result, for given parameter settings, no Schottky barrier is present in this case, and we observe sufficient transmission near the Fermi level, indicating the ohmic nature of the graphene-\ch{MoS2} interface.

To check the reliability of our interface modeling, we integrate the boundary DOS over energy within the band gap of the \ch{MoS2} interior states, and obtain the number of the interfacial states to be about $n \approx 9.4$ $\text{states}$ $\text{nm}^{-1}$. This is close to the full DFT simulation results reported by Chen \textit{et al} \cite{chen2017properties} ranging from 6.3 to 8.3 $\text{states}$ $\text{nm}^{-1}$, and therefore proves the accuracy of our modeling method.
However, the ohmic behavior in our simulation requires doping level one order of magnitude larger than the one reported by experiments. One possible explanation is that we only assume electron hopping between graphene supercells and sulfur atoms immediately next to the interface. By considering interactions of longer range, one can make the overlap states couple better with the conduction band. This could potentially improve transport efficiency and lower the needed doping level to achieve ohmic behavior in our simulation.
Also, although we find that relative perpendicular positions of the two materials hardly affect the transmission efficiency, other boundary configurations including different edge types (armchair or zigzag) and interface roughness could still alter the positions of the overlap states, and in turn change the tunneling current. We leave a systematic examination of the above factors to future work.

\begin{figure}
\includegraphics[width=\linewidth]{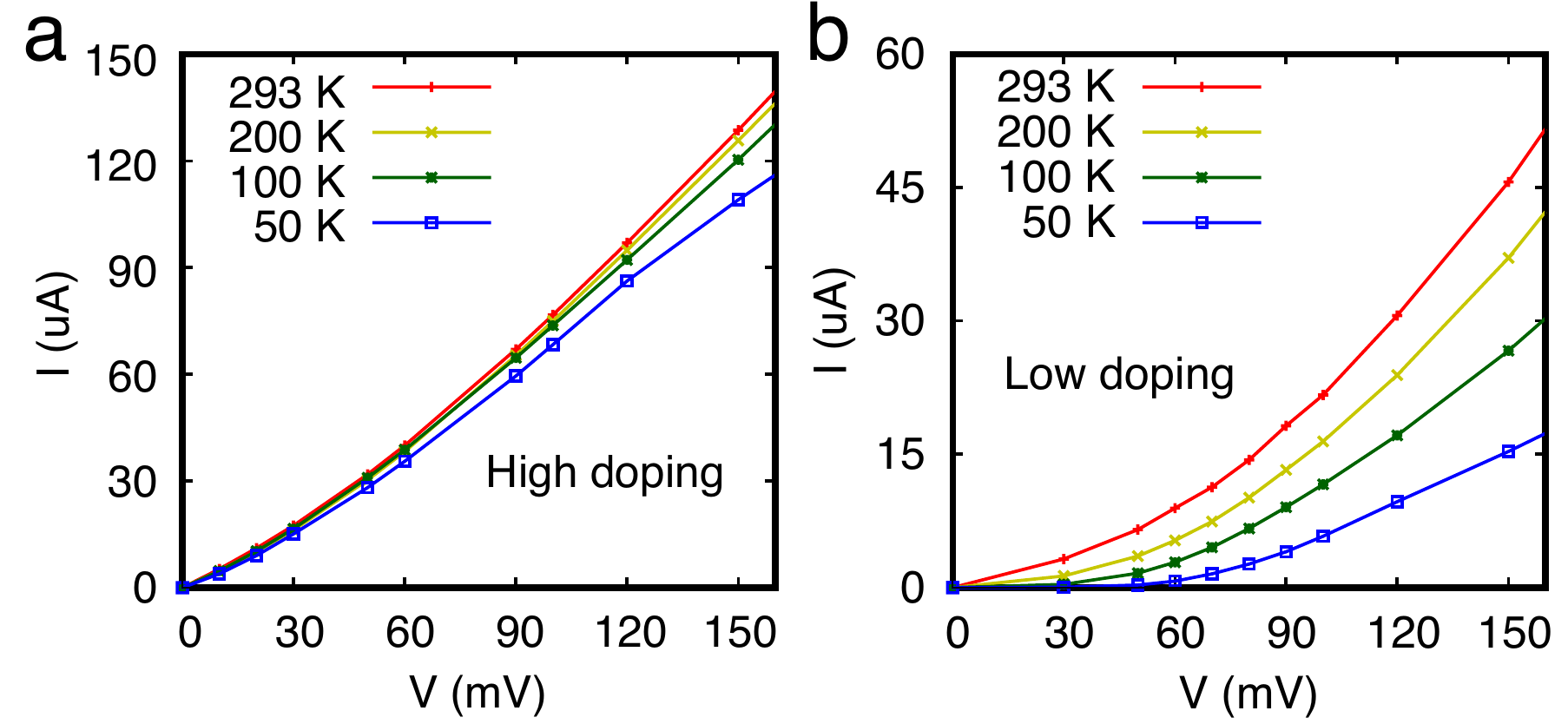}
  \caption{I-V characteristics under different temperatures for \ch{MoS2} doping levels of (a) $ 4\times10^{14}$ $cm^{-2}$, and (b) $2\times10^{14}$ $cm^{-2}$.}
  \label{transport}
\end{figure}

To explore more on the ohmic behavior of graphene-\ch{MoS2} edge contact, we calculate the source-drain currents under different biases and temperatures. The electric current can be calculated as:
\begin{equation}
I(V) = \frac{2e}{h}\int{T(E, V_L, V_R)[f_L(E, V_L) - f_R(E, V_R)]dE}
\end{equation}
where $T(E, V_L, V_R)$ is the transmission coefficient given by equation (\ref{transmission}), $V = V_L - V_R$ is the bias voltage, and $f_{L/R}(E,V_{L/R})$ is the Fermi distribution function of the left/right lead.
In Figure \ref{transport}a, we find that at a high \ch{MoS2} doping level of $4\times10^{14}$ $cm^{-2}$, the I-V curves of the graphene-\ch{MoS2} edge contact show linear characteristics at room temperature, with ohmic behavior maintained down to temperatures at least as low as 50 K. Moreover, we run the same calculation for a \ch{MoS2} doping level of $2\times10^{14}$ $cm^{-2}$, and find non-linear I-V behaviors due to the existence of a large Schottky barrier (see Figure 3b of Supporting Information) at low carrier densities, as shown in Figure \ref{transport}b. This barrier also leads to smaller currents at lower temperatures because reduced thermalization makes it harder for the electrons to go across the interface, which also agrees with experiment. \cite{guimaraes2016atomically}

From Figure \ref{electrostatics}c, we notice that the resonant levels from graphene edge states can assist in the carrier injection across the interface in the range of 0.1 to 0.5 $eV$. One could potentially take advantage of these edge states by adding more n-type doping to graphene and effectively moving down its resonant levels closer to the Fermi level. We verified our prediction by performing simulations with the above changes, and indeed obtained larger transmission values at the Fermi level, as shown in Figure 4 of our Supporting Information. This leads to a lower resistance with a greater source-drain current under the same bias. Therefore, we propose that the usage of an n-doped graphene electrode could be a further improvement to the present edge contact design.

In conclusion, we have developed a computational pipeline to study the electrostatic and transport properties of the 2D graphene-\ch{MoS2} edge contact, and proposed a possible explanation for its ohmic behavior observed in experiments. By applying the custom-built quantum transport simulation scheme, we obtain the charge density profile self-consistently with the electrostatic potential profile of the device. We find that the potential barrier decays fast away from the interface and is thin enough for the electron to tunnel through efficiently. Our results are consistent with both analytical Thomas-Fermi screening theory and the experimental measurements. Because our methods can be scaled effectively to large systems, but maintain the fidelity of {\em ab initio} band structures, they can be used to efficiently predict the electrostatic and transport properties for nanostructures, including those with complex geometries. These findings could have broad implications in the design and fabrication of metal-semiconductor junction for realizing low-resistance contacts.

\begin{acknowledgement}

The authors thank Jiwoong Park and Saptarshi Das for discussions and for comments on an earlier draft of the paper. Research at Argonne is supported by DOE Office of Science, Basic Energy Sciences, Materials Science and Engineering. 

\end{acknowledgement}

\begin{suppinfo}

The supporting information is available free of charge. 
\begin{itemize}
  \item Transport simulation pipeline
  \item Band structure calculation and Wannierization
  \item Self-consistent iteration scheme
  \item Effects of different parameters on the simulation results
  \item Derivation of Thomas-Fermi approximation for quasi-1D systems
\end{itemize}

\end{suppinfo}

\bibliography{edge_contact.bib}

\newpage

\newpage

\newpage

\newpage

\end{document}



\subsubsection{Transport simulation pipeline}
\begin{figure}
\begin{center}
\includegraphics[width=0.9\linewidth]{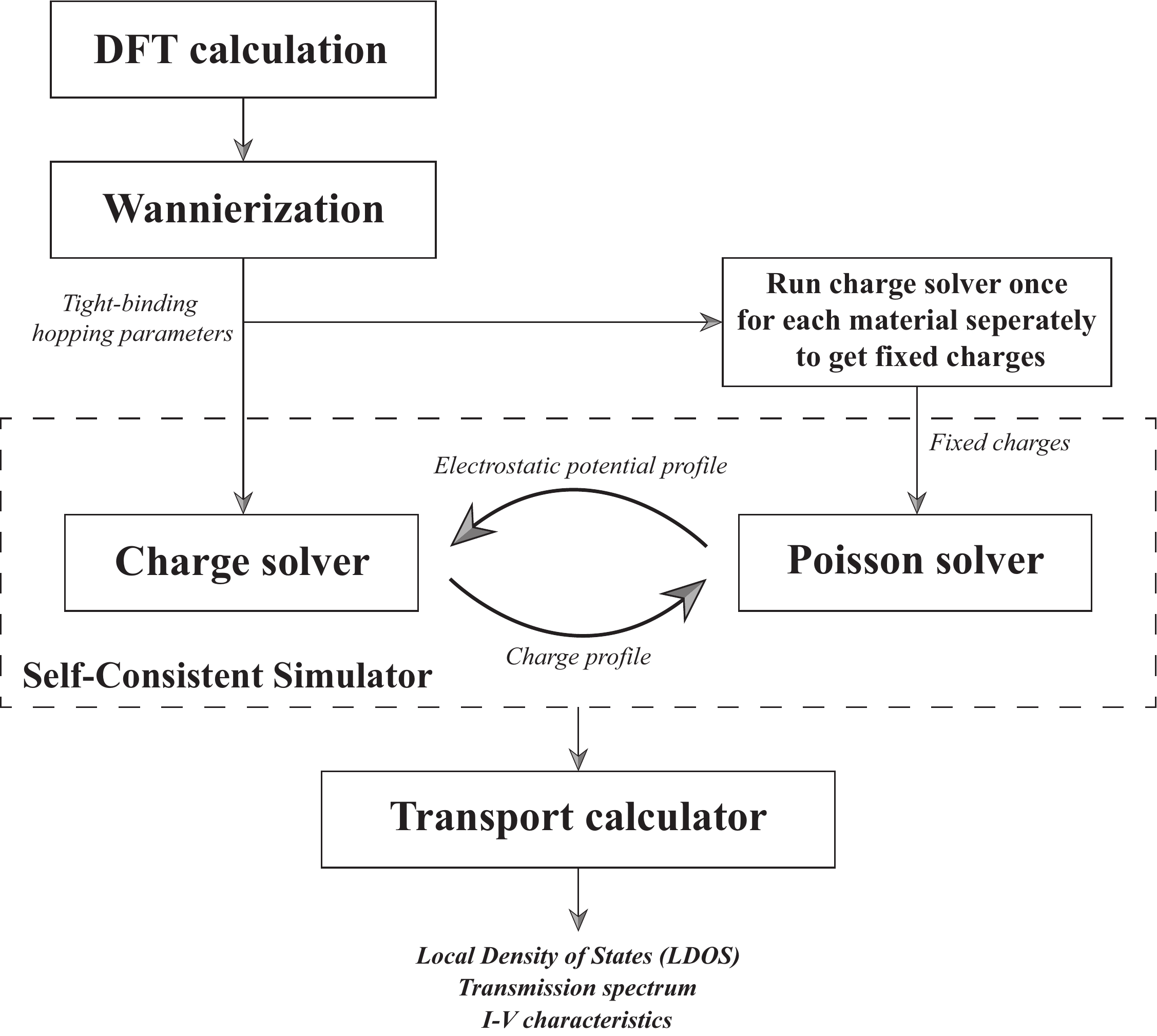}
\end{center}
  \caption{Schematic plot of our simulation pipeline. \textbf{Bold} indicates the main procedures. \textit{Italic} shows the flowing data between different solvers. And \textbf{\textit{bold italic}} at the bottom represents the output.}
  \label{pipeline}
\end{figure}
Figure \ref{pipeline} depicts the block diagram of our simulation pipeline.
In order to accurately describe the electronic structures of considered materials, we first use density-functional-theory (DFT) framework {\sc Quantum ESPRESSO} \cite{giannozzi2009quantum} to calculate their band structures. We then extract tight-binding hopping parameters by the Wannier technique as implemented in the code {\tt wannier90} \cite{mostofi2008wannier90}.
Those parameters allow us to self-consistently solve the electrostatics of the edge contact system using our custom-built software {\tt swan} \cite{wushi_dong_2019_2553787} based on the Keldysh Non-equilibrium Green's Function formalism.
To use our software, we first run the charge solver once separately for each material in order to obtain their fixed charges. Next, using the obtained tight-binding parameters and fixed charges as inputs, we run self-consistent simulations for each bias voltage. After the simulation achieves convergence, we obtain the charge and electrostatic potential profile. Finally, we calculate the local density of states and transmission spectrum using the Landauer-Buttiker formalism, and plot the tunneling currents with biases to generate the I-V characteristic for our simulated edge contact device.

\subsubsection{Band structure calculation and Wannierization}
We performed the DFT calculations as implemented in the package {\sc Quantum ESPRESSO}. We used a plane wave basis set, ultrasoft pseudopotential, and Perdew-Burke-Ernzerhof (PBE) generalized gradient approximation (GGA) exchange-correlation functional, which produces a band gap very close to the experimental one ($E_{g, exp} = 1.8 \text{ eV}$) in the trigonal-prismatic form of single-layer \ch{MoS2} \cite{liu2011performance}. The plane wave cutoff was 100 Ry for wavefunctions and 400 Ry for the charge density. A 15 {\AA} interlayer distance was used to eliminate interlayer interaction. The momentum space was sampled on a $36\times36\times1$ Monkhorst-Pack k-point grid for graphene and $25\times25\times1$ for \ch{MoS2}. Spin-orbit coupling is neglected. The simulated cell is optimized until the atomic forces decrease to values less than $10^{-3}$ {a.u.}. The convergence criterion is set to less than $10^{-6}$ eV total energy difference between two subsequent iterations.

We then used the {\tt wannier90} code\cite{mostofi2008wannier90} to determine the Maximally Localized Wannier Function basis for extracting the tight-binding hopping parameters in the system Hamiltonian. 
Instead of using all the orbitals in the unit cell, we choose only those contributing to the Density of States (DOS) near the Fermi level for Wannier projections. This can minimize the size of matrices used in our calculation and further reduce computational cost. 
We use one atomic $p_z$ orbital for each carbon atom to reproduce the Dirac cones of graphene. In order to capture the seven highest valence bands and four lowest conduction bands of single-layer \ch{MoS2}, we use all three $p$-like Wannier functions centered on each sulfur atom and all five $d$-like projections on each molybdenum.
We included tight-binding parameters up to three orders of nearest neighbors to reproduce the band structure.

For input files of both softwares, please download them via the link below
\newline \url{https://www.dropbox.com/s/0pwfrvmidyigfp7/Input_files.zip?dl=0}.

\subsubsection{Self-consistent iteration scheme}
For an accurate electron transport calculation, it is essential to solve the coupled system of charge and Poisson equations self-consistently.
To evaluate the charge density, we use the Nonequilibrium Green's Function technique based on the Keldysh formalism. We solve the following steady-state kinetic equations to obtain the Green's functions for each electron injection energy E and transverse momentum $k_y$:
\begin{equation}
\sum_{l} \{ [ E - V_i ] \delta_{il} -  H_{il}(k_y) - \Sigma_{il}^{R}(E, k_y)\} G^R_{lj}(E, k_y) = \delta_{ij}
\end{equation}
\begin{equation}
G^{\gtrless}_{ij}(E, k_y) = \sum_{l_1, l_2} G^R_{il_1}(E, k_y) \cdot \Sigma_{l_1l_2}^{\gtrless}(E, k_y) \cdot G^A_{l_2j}(E, k_y)
\end{equation}
The indices $l$, $i$, and $j$ refer to the atomic sites. $H$ is the Hamiltonian of the system. $V_i$ is the self-consistent electrostatic potential at lattice site $i$. The retarded ($G^R$), advanced ($G^A$), greater ($G^>$), and lesser ($G^<$) Green's functions, as well as the boundary self-energies $\Sigma^{R, \gtrless}$, all depend on the electron energy $E$ and momentum $k_y$.
The required elements for calculating the Green's functions are the Hamiltonian and the contact self-energies. Diagonal elements of the Hamiltonian are the on-site energies plus the electrostatic potential energies, which can be obtained self-consistently from the solution of the Poisson equation. The off-diagonal elements represent the hopping strength between neighboring sites. We used the decimation technique \cite{sancho1984quick, sancho1985highly} to calculate the retarded self-energies in order to include the two semi-infinite leads on the left and right respectively.
We also used the recursive Green's function technique to efficiently compute the Green's functions by taking advantage of the sparse nature of Hamiltonian matrices in our calculation \cite{svizhenko2002two}.
The Green's functions and the self-energies are sampled on an energy grid with step sizes comparable to thermal smearing $kT$, where $k$ is the Boltzmann constant and $T$ is temperature. This yields approximately $10^3$ grid points in the energy range considered for transport.
We use 32 $k_y$ points in the irreducible Brillouin zone to represent the periodic y-direction.
The diagonal elements of the lesser Green's function correspond to the spectrum of carrier occupation at a given energy E. So the total electron density at site $i$ is given by
\begin{equation}
n_i = -\frac{2\mathrm{i}}{A\Delta x}\sum_{k_y}\int\frac{dE}{2\pi}G^<_{ii}(k_y, E)
\end{equation}

We then input the charge densities into the following Poisson equation and use a non-linear Newton-Raphson scheme to solve for the electrostatic potential. The Poisson equation for our system is
\begin{equation}
\nabla^2V = -\frac{1}{\epsilon}(-n + N)
\label{poisson}
\end{equation}
where $V$ is the electrostatic potential, $n$ is the electron density, $N$ is the fixed charge density, and $\epsilon$ is the permittivity. In our simulation, we choose the permittivity to be $\epsilon_{graphene} = 6.9$ for monolayer graphene \cite{fang2016microscopic} and $\epsilon_{\ch{MoS2}} = 4.0$ for monolayer \ch{MoS2} \cite{chen2015probing} based on literature results.
Due to the exponential dependence of the carrier concentration on the electrostatic potential, a small potential variation can cause large variation in the carrier concentration and makes it difficult for the Newton-Raphson method to converge. A non-linear scheme takes such an exponential dependence into account and can lead to faster convergence when a linear method cannot even converge at all. Still, we have to use a damping factor between two consecutive Poisson iterations to achieve convergence. A larger damping factor is required for lower temperatures due to small $kT$ values. For example, at the simulation temperature of $50$ K, a damping factor as large as $0.98$ has to be used.

\begin{figure}
\begin{center}
\includegraphics[width=0.7\linewidth]{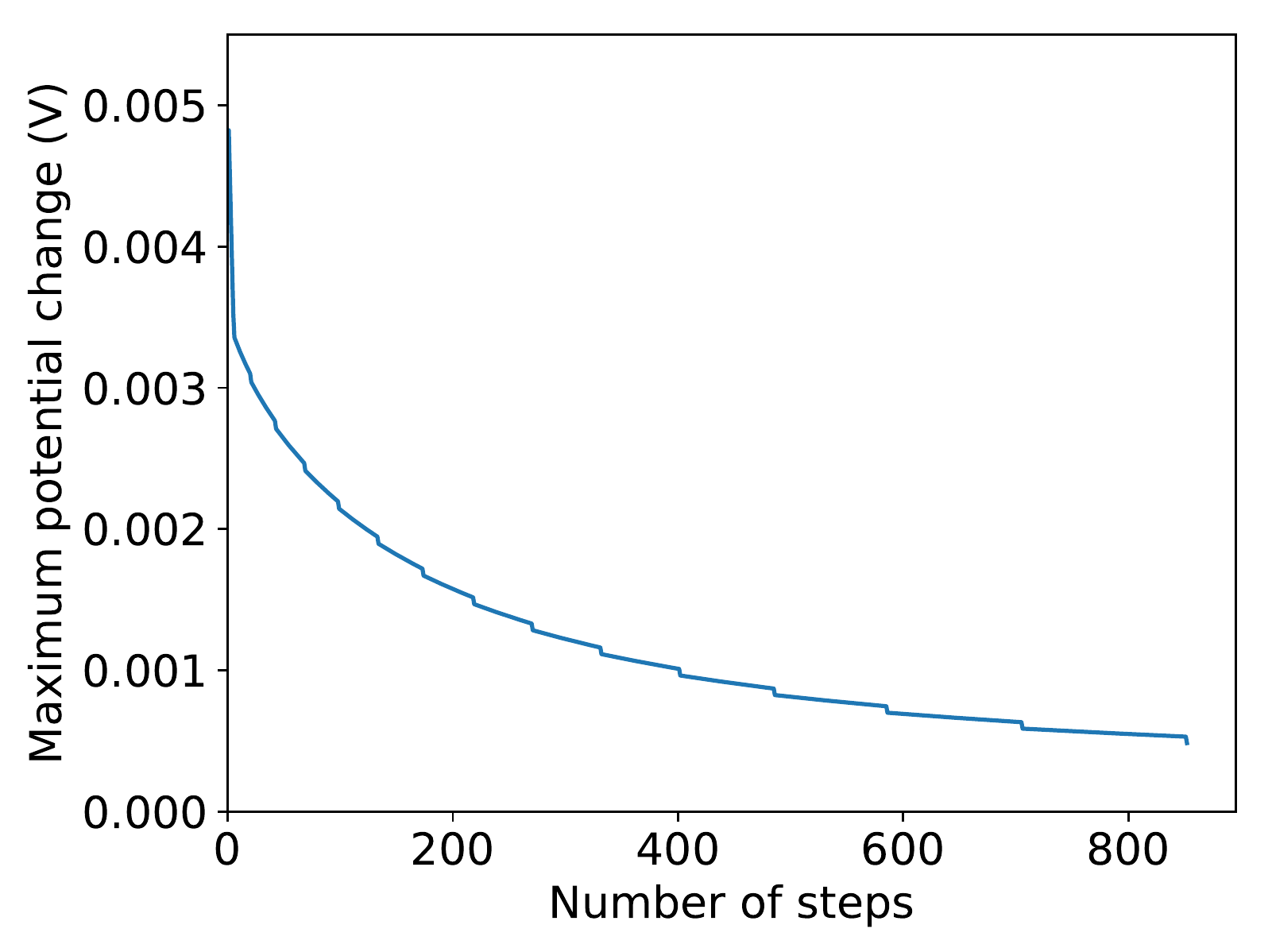}
\end{center}
  \caption{Electrostatic potential converges with the number of run steps in a typical self-consistent simulation.}
  \label{convergence}
\end{figure}

The coupled system of charge and Poisson equations can also be solved by iteration with appropriate numerical damping, which terminates if a convergence criterion is satisfied. In our calculation, the convergence criterion of the self-consistent iterations between two subsequent steps is set to a less than $5\times10^{-4}$ relative difference (with respect to the L1 norm) in the electrostatic potential with the charge density obtained from the previous step. We show the convergence of a typical self-consistent simulation in Figure \ref{convergence}. We can see that our non-linear scheme does a good job at achieving convergence fast within 1000 steps, which usually takes less than 2 hours. After obtaining the converged electrostatic potential, we can evaluate the transport properties based on the Landauer-Buttiker formalism.

\subsubsection{Effects of different parameters on the simulation results}
Here, we show the effects of different model parameters the simulation results, including the local density of states (LDOS) and transmission spectrum. This also demonstrates the robustness of our model. We study below the following four parameters: \ch{MoS2} doping, graphene doping, interfacial hopping strength, and temperature.
The LDOS is evaluated using the the equation:
\begin{equation}
LDOS(E) = \frac{1}{2\pi}[G^r(\Gamma_L + \Gamma_R)G^a]
\end{equation}

\begin{figure}
\begin{center}
\includegraphics[width=0.45\linewidth]{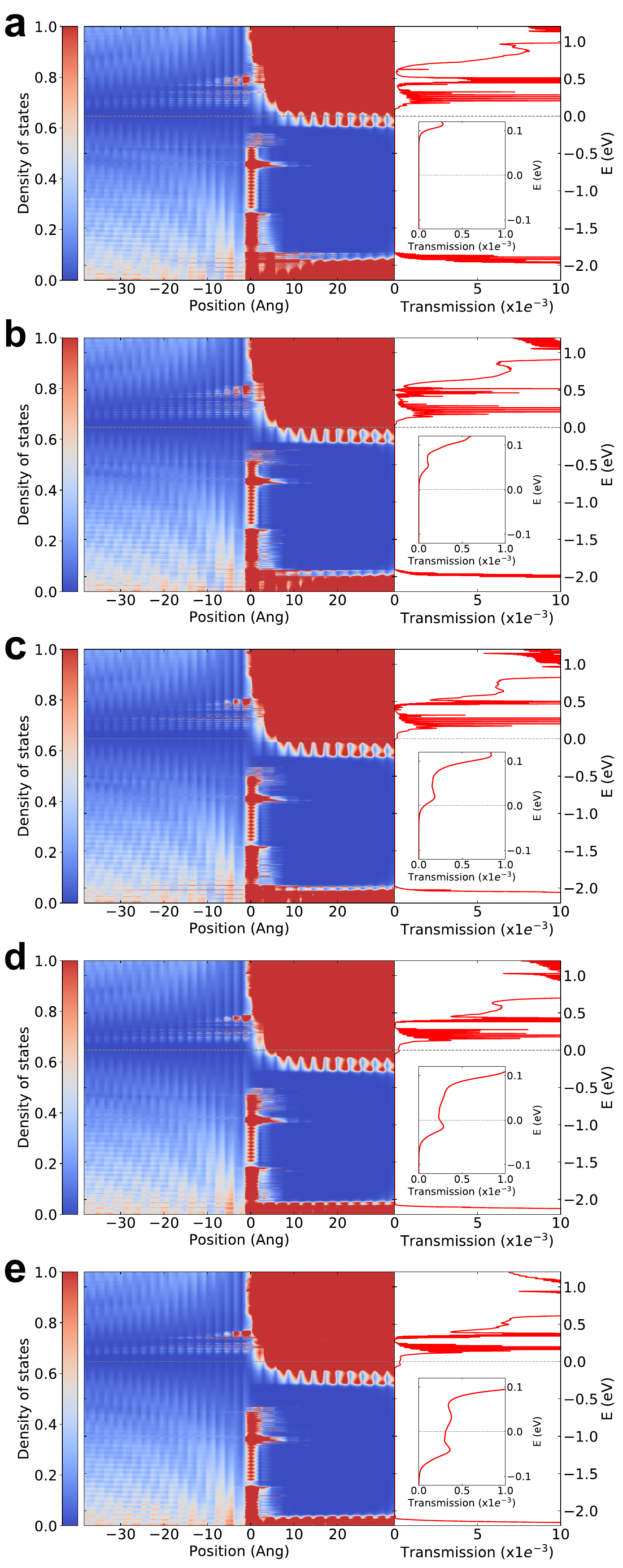}
\end{center}
  \caption{LDOS and transmission spectrum for different \ch{MoS2} doping levels: (a) $1\times10^{14}$ $cm^{-2}$ (b) $2\times10^{14}$ $cm^{-2}$ (c) $3\times10^{14}$ $cm^{-2}$ (d) $4\times10^{14}$ $cm^{-2}$ (e) $5\times10^{14}$ $cm^{-2}$. Inset: Transmission spectrum near the Fermi level at E = 0. ($t_0 = -1.0$ $eV$, $T = 293$ $K$)}
  \label{dopings}
\end{figure}

\paragraph{\ch{MoS2} doping}
We match the doping level induced by gate voltages by adjusting the Fermi level of \ch{MoS2}. For the device used in our manuscript, we used a \ch{MoS2} doping of $4\times10^{14}$ $cm^{-2}$. Here in Figure \ref{dopings}, we compare the LDOS and transmission spectrum for several more doping levels of \ch{MoS2} for room temperature, zero bias and hopping strength of $t_0 = -1.0$ $eV$. We find that the main effect of the \ch{MoS2} doping concentration is shifting the bands of \ch{MoS2}. A higher doping level makes its conduction band minimum move downward, and leads to a larger transmission around the Fermi level.

\begin{figure}
\includegraphics[width=0.6\linewidth]{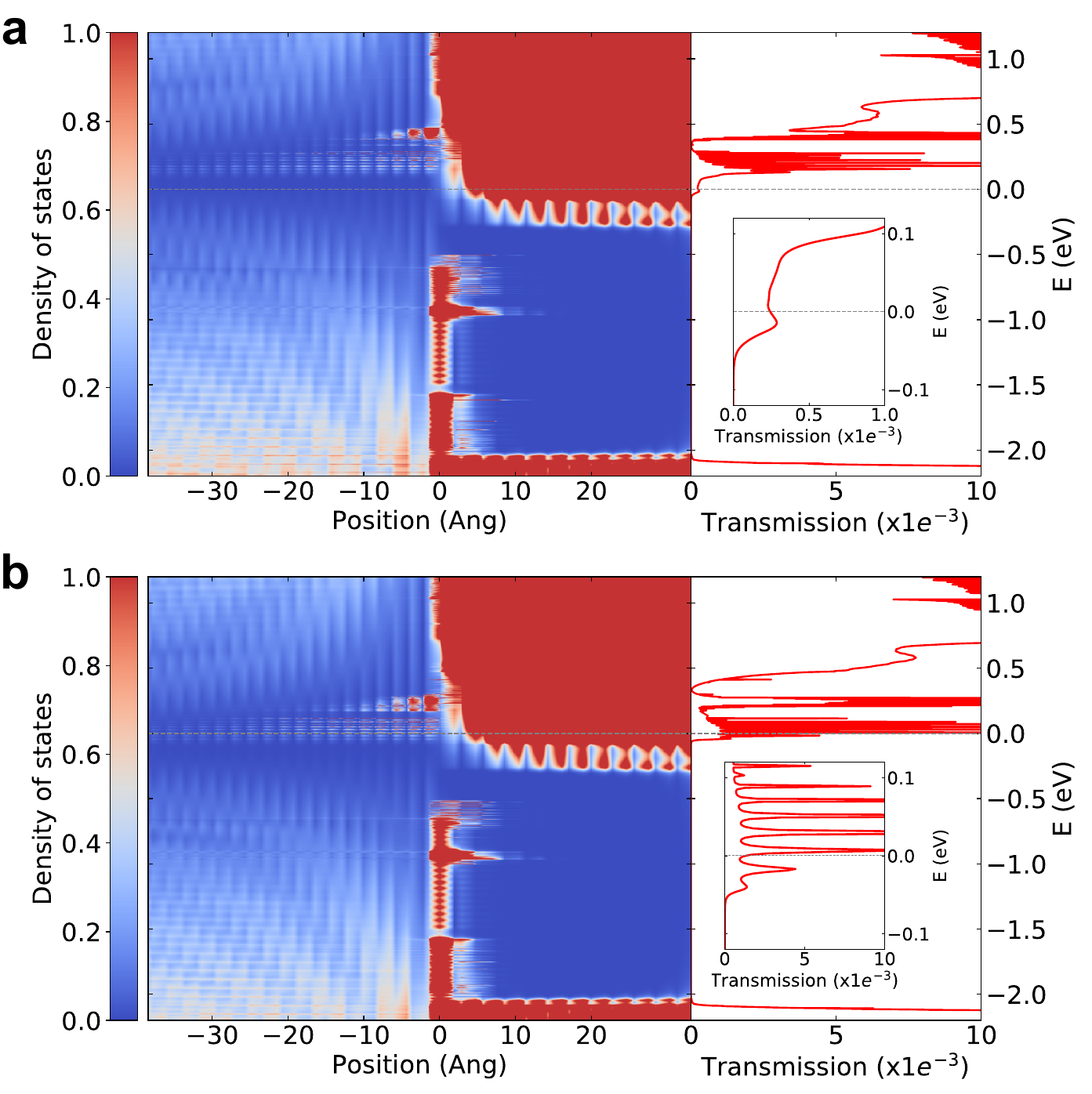}
  \caption{LDOS and transmission spectrum for p-doped and n-doped graphene leads: (a) p-doped (b) n-doped. Inset: Transmission spectrum near the Fermi level at $E = 0$. Notice the transmission scale used in (b) is an order of magnitude larger than that used in (a). (\ch{MoS2} doping $= 4\times10^{14}$ $cm^{-2}$, $t_0 = -1.0$ $eV$, $T = 293$ $K$)}
  \label{gr_doping}
\end{figure}

\paragraph{Graphene doping}
The Fermi level of graphene is lowered to account for an unavoidable p-type doping of approximately $10^{12}$ $cm^{-2}$ resulted from the manufacturing process in the lab. In our paper, we suggest that the usage of n-doped graphene could potentially improve transport efficiency. The related simulation results are shown in Figure \ref{gr_doping}. From the plots, we can see that the transmission in Figure \ref{gr_doping}b for the edge contact device using an n-doped graphene lead is significantly larger than the one in Figure \ref{gr_doping}a using a p-doped graphene lead, especially around the Fermi level. The reason is that the edge states of n-doped graphene are closer to the Fermi level compared to p-doped graphene. This helps the electrons near that energy range to tunnel through and leads to a lager transmission at the Fermi level.

\begin{figure}
\begin{center}
\includegraphics[width=0.5\linewidth]{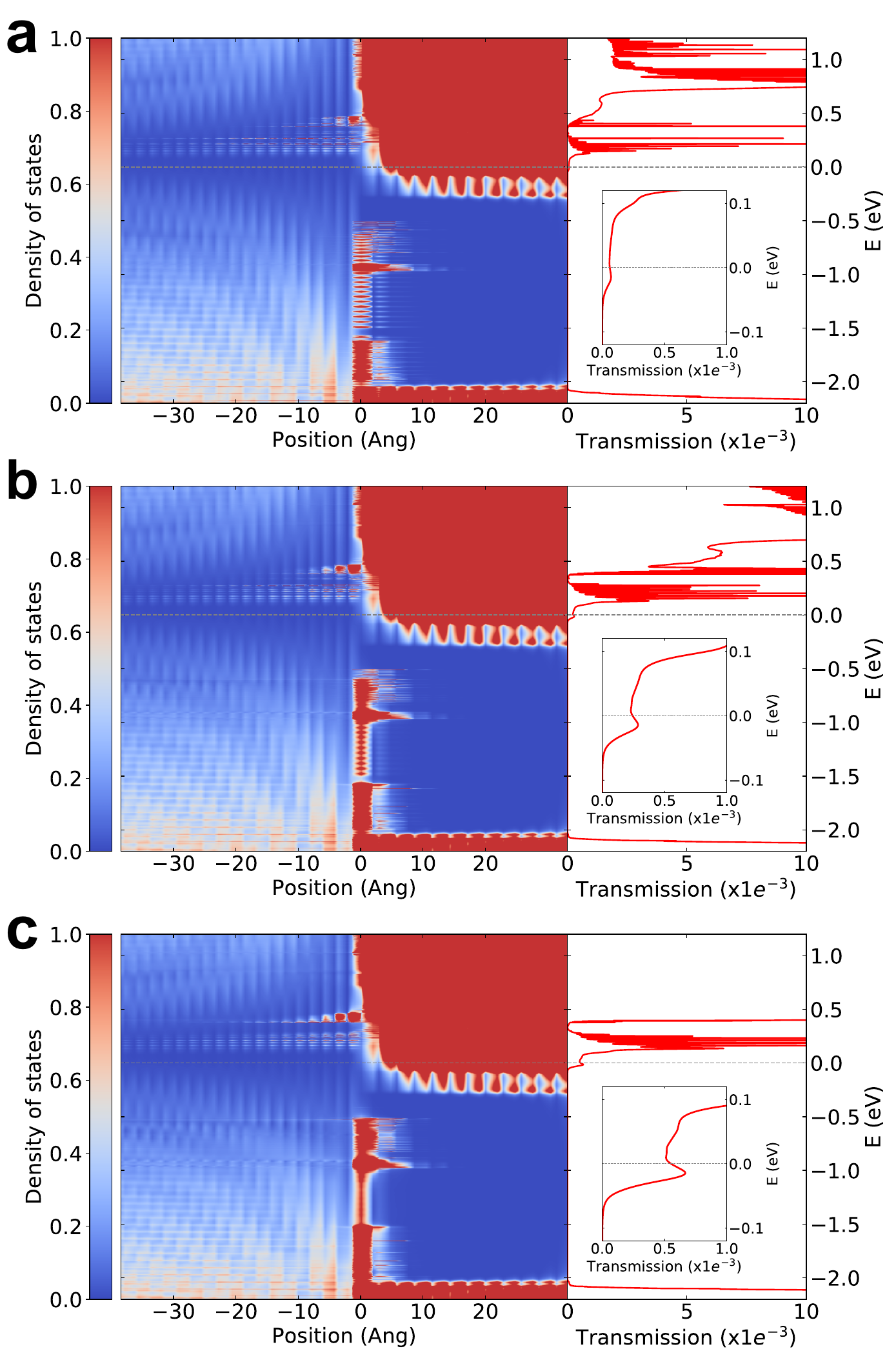}
\end{center}
  \caption{LDOS and transmission spectrum for different interface hopping strengths $t_0$: (a) $t_0 = -0.5$ $eV$ (b) $t_0 = -1.0$ $eV$ (c) $t_0 = -1.5$ $eV$. Inset: Transmission spectrum near the Fermi level at E = 0. (\ch{MoS2} doping $= 4\times10^{14}$ $cm^{-2}$, $T = 293$ $K$)}
  \label{hoppings}
\end{figure}

\paragraph{Interfacial hopping strength}
We consider interfacial interactions only between the $p_z$-like WFs of graphene edge carbon atoms and the $p_x$, $p_y$-like WFs of \ch{MoS2} edge sulfur atoms. The hopping parameter is modeled by an exponential dependence on the distance as $t = t_0e^{-(r - {r_0})}$, where $r$ is the distance between two MLWF centers and $t_0$ is the interaction strength at a distance of $r_0 = 1.5$ {\AA}, which is assumed to be the shortest distance between the graphene and the \ch{MoS2} region. We manually set $t_0$ to be comparable to the strength of the covalent bonds formed at the interface. Comparison of the number of interfacial states in our simulation with the first-principles one proves the accuracy of our interface modeling method.
From Figure \ref{hoppings}, we can see that interfacial hopping strength mainly affects the magnitude of the transmission. Larger interfacial hopping strengths result in larger transmission coefficients.

\begin{figure}[H]
\begin{center}
\includegraphics[width=0.5\linewidth]{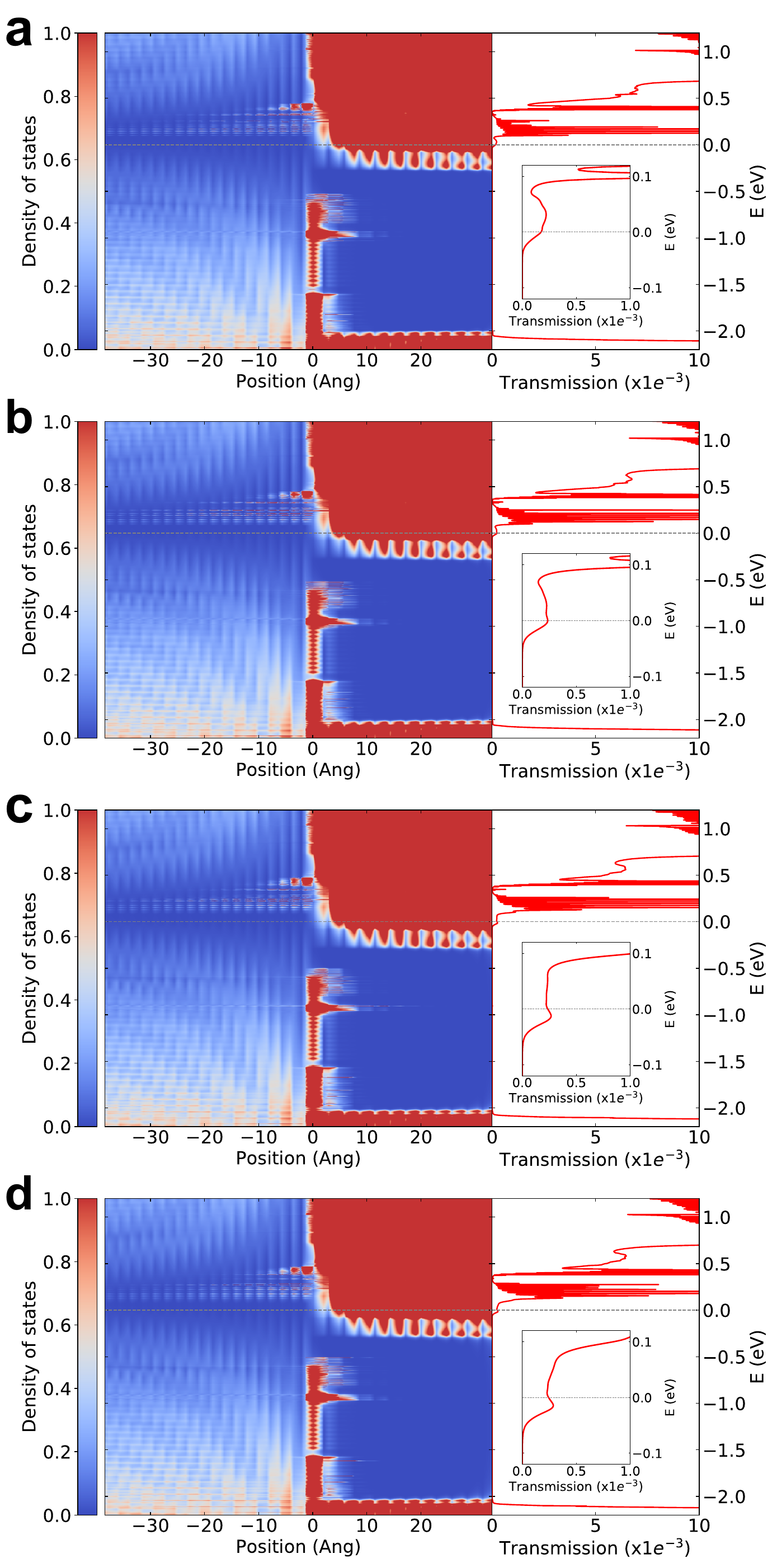}
\end{center}
  \caption{LDOS and transmission spectrum for different temperatures $T$: (a) $T = 50$ $K$ (b) $T = 100$ $K$ (c) $T = 200$ $K$ (d) $T = 293$ $K$. Inset: Transmission spectrum near the Fermi level at E = 0. (\ch{MoS2} doping $= 4\times10^{14}$ $cm^{-2}$, $t_0 = -1.0$ $eV$)}
  \label{temperature}
\end{figure}

\paragraph{Temperature}
In Figure \ref{temperature}, we compare the LDOS and the transmission spectrum at different simulation temperatures. We find that the temperature affects the thermalization of electrons but does not largely alter the transmission spectrum. The reason why we observe smaller currents at lower temperatures is because the decreased thermalization makes it harder for electrons near the Fermi level to overcome the potential barrier. This can also be seen from the calculations of electric current in equation (3) of our manuscript.

\subsubsection{Derivation of Thomas-Fermi approximation for quasi-1D systems}
The derivation begins with the equation for the screened potential energy from an impurity charge distribution $\lambda_i(x)$
\begin{equation}
\nabla^2V(x) = 4\pi e\left[\lambda_i(x) + \lambda_s(x)\right] A
\end{equation}
where $\lambda_s(x)$ is the screening charge and $A$ is the cross section area. We assume $A = 1$ in the following.
The Thomas-Fermi theory approximates the local electron density $n(x)$ as a free-particle system
\begin{equation}
n(x) = \frac{k_F(x)}{\pi}
\end{equation}
where the Fermi wave vector $k_F$ is now a local quantity. It can also be determined by the condition that the chemical potential $\mu$ is independent of position:
\begin{equation}
\frac{k_F^2(x)}{2m} = E_F(x) = \mu - V(x)
\end{equation}
We write the screening charge as the difference between $n(x)$ and the equilibrium charge density $n_0$
\begin{equation}
\lambda_s(x) = -e\left[n(x) - n_0(x)\right]
\end{equation}
and the above approximations result the equation
\begin{equation}
\nabla^2V(x) = 4\pi e\left[\lambda_i(x) + en_0 - en_0\sqrt{1 - \frac{V(x)}{E_F}}\right]
\end{equation}
Assuming $V/E_F << 1$, we can expand the root as $\sqrt{1 - V(x)/E_F} \approx 1 - V/2E_F$ to obtain the equation
\begin{equation}
(\nabla^2 - q_{TF}^2)V(x) = 4\pi e\lambda_i(x)
\end{equation}
\begin{equation}
q_{TF}^2 = \frac{2\pi e^2 n_0}{E_F}
\end{equation}
where $q_{TF}$ is the Thomas-Fermi screening wave vector.
This equation may be solved in 1D Fourier transform space to give
\begin{equation}
V(x) = -4\pi e\int\frac{dq}{2\pi}\frac{\lambda_i(q)}{q^2 + q_{TF}^2}e^{iqx}
\end{equation}
In the case of edge contact, the interfacial trap charge acts as the impurity residing around $x = 0$, and can be represented as $\lambda_i(q) = Q_i$. We can then evaluate the integral and obtain an analytical result
\begin{equation}
V(x) = -\frac{2\pi eQ}{q_{TF}}e^{-q_{TF}|x|}
\end{equation}
The interactions declines rapidly at large distances because of the exponential dependence $e^{-q_{TF}x}$ and thus allows efficient electron tunneling through the interface.


\bibliography{edge_contact_SI}